\documentclass[conference]{IEEEtran}

\IEEEoverridecommandlockouts

\usepackage{cite}
\usepackage{amsmath,amssymb,amsfonts}
\usepackage{graphicx}
\usepackage{xcolor}
\usepackage{chngpage}
\usepackage{multirow}
\usepackage{tabularx}
\usepackage{stmaryrd}
\usepackage{array}
\usepackage{booktabs}
\usepackage{multicol}

\usepackage{enumitem}
\usepackage{url}

\usepackage{booktabs}
\ifCLASSOPTIONcompsoc
  \usepackage[caption=false,font=normalsize,labelfont=sf,textfont=sf]{subfig}
\else
  \usepackage[caption=false,font=footnotesize]{subfig}
\fi

\newcommand{\allClients}{\mathcal{N}}

\newcommand{\clients}{\mathcal{C}}

\newcommand{\corruptedClients}{\eta_C}

\newcommand{\decryptors}{\mathcal{D}}
\newcommand{\dropDecryptors}{\delta_D}
\newcommand{\corruptedDecryptors}{\eta_D}

\newcommand{\inputx}{\boldsymbol{x}_{i}}
\newcommand{\inputxx}{\inputx^\prime}
\newcommand{\indicator}{\boldsymbol{b}_{i}}
\newcommand{\neighbor}{\mathcal{A}_i}
\newcommand{\indices}{\mathcal{B}_{i}}

\newcommand{\allIndices}{\mathcal{K}}

\newcommand{\SAthreshold}{t}
\newcommand{\SSthreshold}{\ell}
\newcommand{\SSshare}{\mathsf{SS.share}}
\newcommand{\SSrecon}{\mathsf{SS.recon}}
\newcommand{\PRG}{\mathsf{PRG}}
\newcommand{\PRF}{\mathsf{PRF}}
\newcommand{\secparam}{\kappa}

\newcommand{\submask}{\boldsymbol{emk}_{u}}
\newcommand{\dropList}{\mathcal{V}}
\newcommand{\maxDropList}{\Delta_{\mathrm{max}}}

\newcommand{\exPart}[1]{\textcolor{red}{#1}}

\newlist{phaseitem}{itemize}{1}
\setlist[phaseitem]{label=, leftmargin=0pt, itemsep=2pt, topsep=0pt, partopsep=0pt}

\newlist{roleitem}{itemize}{1}
\setlist[roleitem]{label=--, leftmargin=1.5em, itemsep=0pt, topsep=0pt, partopsep=0pt}

\newlist{actionitem}{itemize}{1}
\setlist[actionitem]{label=\textbullet, leftmargin=1em, itemsep=0pt, topsep=0pt, partopsep=0pt}

\newcommand{\CenterRow}[2]{
  \dimen0=\ht\strutbox%
  \advance\dimen0\dp\strutbox%
  \multiply\dimen0 by#1%
  \divide\dimen0 by2%
  \advance\dimen0 by-.5\normalbaselineskip%
  \raisebox{-\dimen0}[0pt][0pt]{#2}}

\usepackage{amsthm}

\theoremstyle{plain} 
\newtheorem{thm}{Theorem}

\theoremstyle{definition} 

\newenvironment{sketchproof}{\begin{proof}[Sketch Proof]}{\end{proof}}

\begin{document}

\title{Per-element Secure Aggregation against Data Reconstruction Attacks in Federated Learning
}


\author{\IEEEauthorblockN{Takumi Suimon\IEEEauthorrefmark{1}, Yuki Koizumi\IEEEauthorrefmark{1}, Junji Takemasa\IEEEauthorrefmark{1}, and Toru Hasegawa\IEEEauthorrefmark{2}}
\IEEEauthorblockA{\IEEEauthorrefmark{1}Graduate School of Information Science and Technology, The University of Osaka}
\IEEEauthorblockA{\IEEEauthorrefmark{2}Faculty of Materials for Energy, Shimane University}
}

\maketitle

\begin{abstract}
Federated learning (FL) enables collaborative model training without sharing raw data, but individual model updates may still leak sensitive information.
Secure aggregation (SecAgg) mitigates this risk by allowing the server to access only the sum of client updates, thereby concealing individual contributions.
However, a significant vulnerability has recently attracted increasing attention: when model updates are sparse vectors, a non-zero value contributed by a single client at a given index can be directly revealed in the aggregate, enabling precise data reconstruction attacks.
In this paper, we propose a novel enhancement to SecAgg that reveals aggregated values only at indices with at least $\SAthreshold$ non-zero contributions.
Our mechanism introduces a per-element masking strategy to prevent the exposure of under-contributed elements, while maintaining modularity and compatibility with many existing SecAgg implementations by relying solely on cryptographic primitives already employed in a typical setup.
We integrate this mechanism into Flamingo, a low-round SecAgg protocol, to provide a robust defense against such attacks.
Our analysis and experimental results indicate that the additional computational and communication overhead introduced by our mechanism remains within an acceptable range, supporting the practicality of our approach.
\end{abstract}

\begin{IEEEkeywords}
Federated learning, Secure aggregation, Data reconstruction attacks
\end{IEEEkeywords}



\section{Introduction}
\label{sec:introduction}
Federated learning (FL)~\cite{McMahan2017-sx} is widely adopted for its ability to train models on distributed data while preserving data privacy.
In each training round, clients independently train the received global model on their datasets and send model updates to the server.
Please note that models and updates are represented as vectors, and the terms ``models/updates'' and ``vectors'' are used interchangeably throughout this paper.
The server aggregates these updates to produce a new global model, which is again distributed to clients.
The privacy principle of FL is that only updates are shared under the assumption that these updates do not reveal the raw datasets.
However, numerous studies~\cite{Wang2019-fz, Geiping2020-my} have shown that a motivated server can often reconstruct clients' data from their individual updates.

Secure aggregation (SecAgg)~\cite{Bonawitz2017-vr, Bell2020-vn, Ma2023-fq, Karthikeyan2024-sm, Bell-Clark2024-hc} is a key countermeasure to this privacy risk.
It ensures that the server receives only the aggregate of at least $\SAthreshold$ vectors (where $\SAthreshold \geq 2$), without revealing any individual vector.
When integrated into FL, SecAgg prevents the server from accessing individual updates and mitigates the risk of data reconstruction.

However, a critical vulnerability occurs when the vectors are sparse.
SecAgg hides individual vectors but does not always hide individual elements.
If only one client provides a non-zero value at a given index in the vector, that value can be revealed from the aggregate result.
This loophole poses a serious privacy threat.
Recent studies~\cite{Fowl2022-an, Wen2022-rw, Boenisch2023-ow, Pasquini2022-sb, Zhao2024-wy, Chu2022-bu} propose data reconstruction attacks under SecAgg, where the server maliciously crafts global models so that updates from non-victim clients are zero at specific indices, exposing corresponding elements in the victim clients’ updates.

To counter such attacks, two types of defenses have been proposed: model inconsistency checks and model integrity checks.
However, both approaches rely on client-side model checks, which suffer from fundamental limitations.
First, \textbf{model inconsistency checks}~\cite{Pasquini2022-sb, Sultan2025-hq} attempt to detect attacks by verifying whether all clients receive the same model, which prevents a malicious server from delivering different modified models to different clients.
Unfortunately, this type of defense is not applicable to modern FL, such as personalized FL~\cite{Tan2025-zh} and asynchronous FL~\cite{So2021-ea, Nguyen2022-rv}, where the server inherently distributes different models.
In addition, model inconsistency checks fail to detect attacks where the same modified model is sent to all clients~\cite{Wen2022-rw, Zhao2024-wy}, as no inconsistency arises to be detected.

Second, \textbf{model integrity checks}~\cite{Xu2020-qf, Zhu2024-kl, Guan2025-jt, Garov2023-lu, Wang2025-vs} aim to detect attacks by verifying whether the received model has been maliciously modified.
This class of defenses falls into two main categories, each with drawbacks.
One approach relies on cryptographic verification to ensure that the global model correctly reflects client updates~\cite{Xu2020-qf, Zhu2024-kl, Guan2025-jt}.
While this provides strong security, it relies on costly cryptographic techniques such as zero-knowledge proofs, which significantly increase computational overhead.
Another approach inspects the model for anomalous parameters or structures~\cite{Garov2023-lu, Wang2025-vs}.
While lightweight, in FL settings where clients receive only partial models~\cite{Horvath2021-nl, Wu2024-wh}, it becomes impossible to verify the unseen parts.
These parts are treated as zero elements, which opens the door to data reconstruction attacks.

This paper proposes a fundamental mechanism to prevent the exposure of individual element values unless at least~$\SAthreshold$ non-zero contributions are made at a given index. 
This mechanism inherently nullifies data reconstruction attacks, as it prevents the server from unmasking elements contributed by too few clients.
The key idea is to introduce two types of new vectors: one for counting non-zero contributors and the other for masking model updates.
In the SecAgg procedure, each client masks its model update using both conventional and newly introduced masks; at a given index, the new mask is sent to the server only if the number of contributors exceeds a certain threshold.
Our approach does not rely on client-side model checks and remains simple, thereby avoiding the inherent limitations of existing methods.

The contributions of this paper are as follows.
First, we propose \textit{Per-element SecAgg}, a framework that reveals aggregated values only at indices with at least $\SAthreshold$ non-zero contributions.
Our design does not require any new cryptographic primitives beyond standard SecAgg protocols~\cite{Bonawitz2017-vr, Bell2020-vn, Ma2023-fq}.
Second, we develop a full protocol by extending Flamingo~\cite{Ma2023-fq}, a lightweight SecAgg.
Our protocol ensures that the guarantees of Per-element SecAgg hold even in the presence of a malicious server, user dropouts, and user collusion.
Third, we implement the protocol and evaluate its performance.
The results show that the additional overhead is acceptable, and that the impact of Per-element SecAgg on model performance is minimal, confirming that our protocol is practical.

The rest of the paper is organized as follows: 
Section~\ref{sec:preliminaries} explains the preliminaries. 
Sections~\ref{sec:framework} and~\ref{sec:protocol} describe the design rationale of our Per-element SecAgg framework and the protocol design, respectively.
Section~\ref{sec:analysis} analyzes the cost and the security, and Section~\ref{sec:evaluation} evaluates the performance of the protocol.
Section~\ref{sec:related} summarizes the related work, and Section~\ref{sec:conclusion} concludes the paper.

\begin{table}[t]
  \centering
    \caption{Notation frequently used in this paper.}
  \label{tab:notation}
  \begin{tabular}{p{0.15\linewidth}p{0.75\linewidth}}
    \toprule
    Notation & Description \\
    \midrule
    $\allClients$           &   Set of all users\\
    $\clients, \decryptors$ &   Set of clients and decryptors, respectively \\

    $\dropDecryptors$             &   Dropout rate of decryptors  \\
    $\corruptedClients, \corruptedDecryptors$   &   Colluding rate of clients and decryptors, respectively \\

    $\inputx$               &   Model update vector of client $i \in \clients$ \\
    $\allIndices$           &   All indices in model update vector~$\inputx$ \\

    $\SAthreshold, \SSthreshold$    &   Threshold in SecAgg and secret sharing, respectively \\

    \bottomrule
  \end{tabular}
\end{table}

\section{Preliminaries}
\label{sec:preliminaries}
In this section, we introduce the cryptographic primitives and the Flamingo SecAgg protocol, which are essential for understanding this paper.
Table~\ref{tab:notation} summarizes the notation.

\subsection{Cryptographic Primitives}
\label{subsec:crypto-primitive}
\subsubsection{Diffie-Hellman Key Exchange via PKI}
A user~$i$ generates a private key~$a_i \in \{0,1\}^\secparam$ and publishes the public key~$g^{a_i}$ via a public key infrastructure~(PKI), where~$\secparam$ is a secyrity parameter and~$g$ is a generator of a cyclic group with prime order.
A pair of users~$i,j$ establishes a shared secret~$s_{i,j} = g^{a_i a_i}$ using their private key and the other's public key.

\subsubsection{Pseudorandom Generators}
$\PRG(r)$ is a deterministic function that expands a random seed~$r \in \{0,1\}^\secparam$ into a longer pseudorandom string.
The PRGs used in this paper are secure, meaning their outputs are computationally indistinguishable from truly random strings, as long as the seeds remain secret.

\subsubsection{Pseudorandom Functions}
$\PRF(k,x)$ is a deterministic function that maps an input~$x$ to a pseudorandom string of the same length, using a secret key~$k \in \{0,1\}^\secparam$.
The PRFs used in this paper are secure, meaning their outputs are computationally indistinguishable from truly random strings, as long as the keys remain secret.

\subsubsection{Secret Sharing}
Shamir’s~$(\SSthreshold, L)$ secret sharing scheme distributes a secret~$s$ using two algorithms:
$\SSshare(s, \SSthreshold, L) \rightarrow \{\langle s \rangle_1, \dots, \langle s \rangle_L \}$ splits~$s$ into~$L$ shares, and
$\SSrecon(\{\langle s \rangle_l\}_{l \in \mathcal{L}}) \rightarrow s$ reconstructs~$s$ from any subset with~$|\mathcal{L}| \geq \ell$.
The scheme ensures that fewer than~$\ell$ shares reveal no information about~$s$.

\subsection{Flaminngo SecAgg Protocol}
\label{subsec:flamingo}
SecAgg enables the server to obtain the sum of client inputs while keeping individual inputs private.
We focus on the Flamingo~\cite{Ma2023-fq} protocol because it fits seamlessly into our system model defined in Section~\ref{subsec:system-model}.
A key innovation of Flamingo lies in its highly efficient unmasking phase, which is a bottleneck in \cite{Bonawitz2017-vr}.
Instead of involving all clients, Flamingo offloads this task to decryptors, a small set of users.

\subsubsection{Setup Phase}
\label{subsubsec:flamingo-setup}
The setup phase is executed only once before training begins.
During this phase, each pair of users~$i,j \in \allClients$ establishes a long-term secret~$s_{i,j}$ and a symmetric key~$k_{i,j}$ using Diffie-Hellman key exchange through PKI.
Long-term secrets serve as the basis for generating pairwise masks in subsequent rounds.
In addition, a trusted source of public randomness~$R$~\cite{randomness-beacon} is used to select a small subset of users as decryptors~$\decryptors$ from $\allClients$.

\subsubsection{Report Phase}
\label{subsubsec:flamingo-report}
In each round~$\tau$, each client~$i \in \clients$ uses~$R$ to determine a set of neighbors~$\neighbor \subset \clients$ with whom it will compute pairwise masks.
Using the long-term secrets,~$i$ derives round-specific pairwise seed~$r_{i,j} \leftarrow \PRF(s_{i,j}, \tau)$ for all peers~$j \in \neighbor$.
In addition,~$i$ randomly generates an individual seed~$r_i$.
Using these,~$i$ masks its model update vector~$\inputx$ as follows:
\begin{equation}
\begin{aligned}
    \llbracket \inputx \rrbracket = \inputx 
    + \underbrace{\PRG(r_{i})}_{\text{(individual mask)}} 
    + \underbrace{\sum_{j \in \neighbor} \pm \PRG(r_{i,j})}_{\text{(pairwise mask)}},
\end{aligned}
\end{equation}
where the sign of each pairwise mask is positive if~$i<j$ and negative otherwise.
Although individual masks are essential for ensuring robustness against client dropouts, we assume no dropouts occur in this explanation for simplicity.

To enable unmasking of the individual mask,~$i$ secret-shares~$r_i$ as~$\{\langle r_i \rangle_u\}_{u \in \decryptors} \leftarrow \SSshare(r_i, \SSthreshold, |\decryptors|)$.
Then, it encrypts each share~$\langle r_i \rangle_u$ using the symmetric key~$k_{i,u}$ shared with decryptor~$u$.
That is, a total of~$|\decryptors|$ ciphertexts~$\{\llbracket \langle r_i \rangle_u \rrbracket_{k_{i,u}}\}_{u \in \decryptors}$ are generated.
Finally,~$i$ sends the masked input~$\llbracket \inputx \rrbracket$ along with these ciphertexts to the server.
Although there are~$|\decryptors|$ ciphertexts, each seed and its share are significantly smaller in size than the masked input vector itself, resulting in only marginal communication overhead.

\subsubsection{Unmasking Phase}
\label{subsubsec:flamingo-unmasking}
The server aggregates all the received masked vectors.
Since we assume no client dropouts, the pairwise masks cancel out upon aggregation, yielding:
\begin{equation}
    \sum_{i\in \clients}{\llbracket \inputx \rrbracket} = \sum_{i\in \clients}{\inputx} + \sum_{i\in \clients}{\PRG(r_i)}.
\end{equation}
The goal of this phase is to remove individual masks so that the server can recover the plaintext~$\sum_{i\in \clients}{\inputx}$.

To this end, the server sends the ciphertexts of seed shares to their corresponding decryptors.
Specifically, each decryptor~$u \in \decryptors$ receives a set of ciphertexts~$\{\llbracket \langle r_i \rangle_u \rrbracket_{k_{i,u}}\}_{i \in \clients}$, which it decrypts using the symmetric keys~$\{k_{i,u}\}_{i\in \clients}$.
The plaintext seed shares are then returned to the server.

The server can reconstruct each seed~$r_i$ only if it collects at least~$\SSthreshold$ shares from a subset of decryptors~$\decryptors_1 \subseteq \decryptors$, i.e.,~$r_i \leftarrow \SSrecon(\{\langle r_i \rangle_u \}_{u \in \decryptors_1})$, where~$|\decryptors_1| \geq \SSthreshold$.
Finally, the server regenerates the individual masks by applying PRGs to each reconstructed seed and subtracts the total mask~$\sum_{i\in \clients}{\PRG(r_i)}$ from the aggregated vector to obtain the plaintext sum~$\sum_{i\in \clients}{\inputx}$.

\section{Per-element SecAgg Framework}
\label{sec:framework}
This section defines the system and threat models along with our design goals.
Based on them, we propose the Per-element SecAgg framework against data reconstruction attacks.

\subsection{System Model}
\label{subsec:system-model}
We consider a star topology with a single server and a set of ~$\allClients$ users.
Following prior work on Flamingo~\cite{Ma2023-fq}, Willow~\cite{Bell-Clark2024-hc} and OPA~\cite{Karthikeyan2024-sm}, we assume two roles among users: clients and decryptors. 
In each round~$\tau$, a set of clients~$\clients \subset \allClients$ is selected at random based on a public source of randomness.
Each client~$i \in \clients$ receives the global model from the server, trains it on its local dataset, and returns an update vector~$\inputx \in \mathbb{R}^{|\allIndices|}$.
Separately, a set of decryptors~$\decryptors \subset \allClients$ is randomly selected using the same public randomness. 
Decryptors assist in the SecAgg procedure but do not participate in model training.
The server receives only the aggregated sum of the update vectors~$\sum_{i\in \clients}{\inputx}$ via the SecAgg protocol, without learning any individual update.

Some users may drop out during the protocol due to unstable network conditions or battery limitations.
We denote the dropout rate of decryptors in each round by~$\dropDecryptors$.
We do not explicitly define the dropout rate of clients, as their dropout does not affect our Per-element SecAgg mechanism.

\subsection{Threat Model}
\label{subsec:threat-model}
We assume that the server behaves as a malicious adversary.
It is allowed to arbitrarily deviate from the SecAgg protocol, including manipulation of protocol messages.
In addition, the server can freely modify the global model distributed to clients in each round.
The server may also collude with up to a fraction~$\corruptedClients$ of clients and a fraction~$\corruptedDecryptors$ of decryptors, from whom it can obtain internal protocol information such as secret keys or secret shares.
As with Flamingo, we assume that~$\dropDecryptors + \corruptedDecryptors < 1/3$ and treat dropped decryptors and colluding decryptors as disjoint sets.
The adversary’s goal is to reconstruct the datasets of honest clients under SecAgg by launching sparse update-based attacks~\cite{Fowl2022-an, Wen2022-rw, Boenisch2023-ow, Pasquini2022-sb, Zhao2024-wy, Chu2022-bu}.

\subsection{Design Goals}
\label{subsec:design-goals}

Our primary goal is to prevent data reconstruction attacks by simultaneously satisfying the following properties (P1)--(P4).

\noindent
\textbf{(P1) Per-element Threshold Aggregation. }
The server learns only the aggregated values of elements that are contributed by at least~$\SAthreshold$ honest clients.
This property is the core of our work.

\noindent
\textbf{(P2) Security against a malicious server. }
Even if the server behaves maliciously, (P1) is still guaranteed.
Specifically, this property considers the case where the malicious server sends forged information to honest decryptors.

\noindent
\textbf{(P3) Dropout Tolerance. }
Even if up to~$\lfloor \dropDecryptors \decryptors \rfloor$ decryptors drop out, the protocol does not abort, and (P1) remains guaranteed.
Tolerance to client dropouts is also required, but this can be addressed by existing SecAgg protocols~\cite{Bonawitz2017-vr, Bell2020-vn, Ma2023-fq} without affecting our framework.
Therefore, the rest of this paper does not elaborate on recovery from client dropouts.

\noindent
\textbf{(P4) Collusion resilience. }
Even if the server colludes with up to~$\lfloor \corruptedClients \clients \rfloor$ clients and up to~$\lfloor \corruptedClients \decryptors \rfloor$ decryptors, (P1) is still guaranteed.

On the other hand, the following properties are not part of our goals.

\noindent
\textbf{Prevention of other attacks.}
We focus on defending against data reconstruction attacks, the most critical privacy threat in FL.
Therefore, other attacks, such as property inference attacks~\cite{Pasquini2022-sb} and label inference attacks~\cite{Wang2024-yx}, are out of scope.

\noindent
\textbf{Privacy of index information. }
Our approach reveals to the server and decryptors the indices where each client contributes non-zero values, as is commonly done for sparse FL~\cite{Ergun2022-mn}.
While such exposure may raise privacy concerns, we do not attempt to hide them in our design, as it poses little threat. 
A detailed discussion is deferred to Section~\ref{subsec:privacy-analysis}.




\subsection{Core Mechanism and Rationale}
\label{subsec:core-mechanism}

\begin{figure*}[t]
   \centering
   \includegraphics[width=0.85\linewidth]{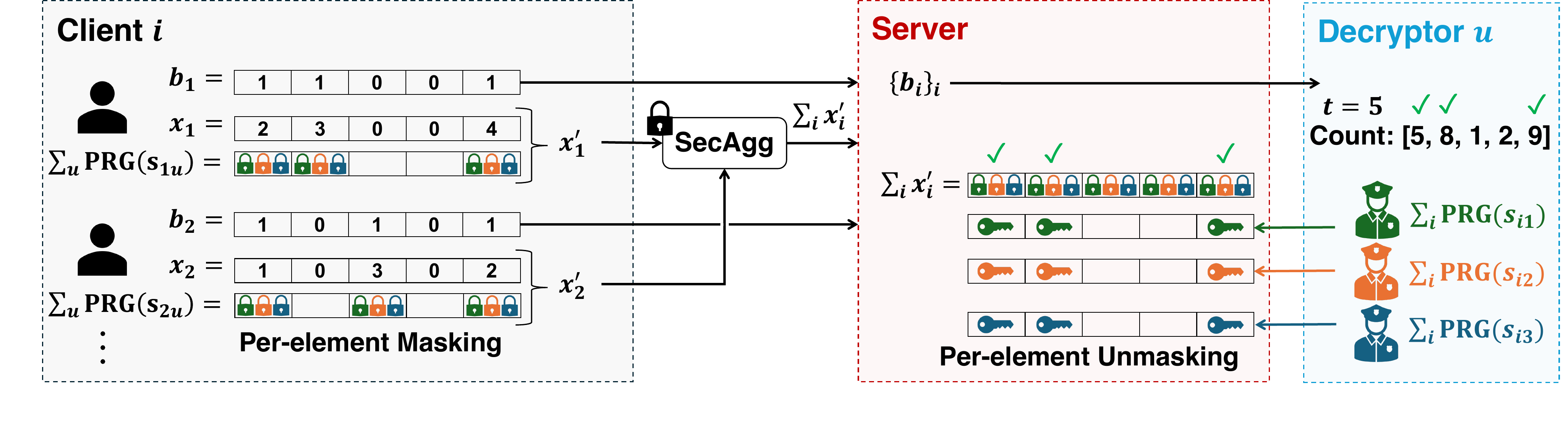}
    \caption{Overview of Per-element SecAgg framework.}
 \label{fig:core-idea}
\end{figure*}

We introduce a lightweight mechanism that extends the SecAgg process to support Per-element SecAgg.
This mechanism is built solely on standard cryptographic primitives (Section~\ref{subsec:crypto-primitive}).
The key idea is to enable decryptors to control which elements are unmasked based on how many clients contribute non-zero values at each index.

The protocol proceeds as follows (illustrated in Fig.~\ref{fig:core-idea}):
\begin{itemize}
    \item Each client adds additional pairwise masks---shared with decryptors---to its local update vector, but only at the indices corresponding to non-zero elements.
    \item After collecting the masked vectors, the server counts how many clients contributed non-zero values at each index and forwards this information to the decryptors.
    \item For each index that exceeds threshold~$\SAthreshold$, the decryptors return the corresponding pairwise mask, enabling the server to unmask the element of the aggregated vectors.
\end{itemize}

This overall flow is designed to satisfy our key security properties (P1) and (P2).
In what follows, we explain the rationale behind each of the three components of the mechanism:
1) per-element contributor counting,
2) per-element masking, and
3) per-element unmasking.
The overall protocol, along with the designs addressing (P3) and (P4), is presented in Sections~\ref{sec:protocol} and~\ref{sec:analysis}, respectively.

\subsubsection{Per-element Contributor Counting}
\label{subsubsec:per-element-counting}
To realize (P1), the decryptors must determine which indices have received at least~$\SAthreshold$ non-zero contributions.
To this end, each client~$i\in \clients$ generates a binary indicator vector~$\indicator$, where:
\begin{equation}
    \label{eq:indicator}
    \indicator[k] =
    \begin{cases}
        1 & \text{if} \; \inputx[k] \neq 0 \\
        0 & \text{otherwise,}
    \end{cases}
\end{equation}
for all~$k \in \allIndices$.
This vector flags the indices where the client's local update vector~$\inputx$ has non-zero values.
Clients send their indicator vectors to the server, which collects and forwards~$\{\indicator\}_{i \in \clients}$ to the decryptors, in accordance with the star topology of the network.
This enables the decryptors to compute, for each index, the number of clients that have contributed non-zero values, and decide whether to unmask the corresponding element accordingly.

\subsubsection{Per-element Masking}
\label{subsubsec:per-element-masking}
For (P1), the server forwards the set of indicator vectors~$\{\indicator\}_{i \in \clients}$ to the decryptors.
However, a malicious server may forge these vectors, falsely claiming that an index~$k$ has received at least~$\SAthreshold$ non-zero contributions.
This may cause honest decryptors to return masks for elements that do not genuinely satisfy the threshold condition, allowing the server to unmask them and thus violating~(P2).

A straightforward solution would be to make each~$\indicator$ verifiable using cryptographic tools.
However, such an approach introduces additional cryptographic primitives and complicates the protocol.
Instead, we adopt a simpler and more robust design: we tolerate the possibility of forged indicators, but ensure that any manipulation by the server cannot result in successful unmasking.

To this end, each client adds pairwise masks with the decryptors to its local update~$\inputx$, but only at indices with non-zero values.
As a result, the total mask at each index reflects only the contributions from clients with actual non-zero values at that index.
Any mask derived from forged indicators fails to match this sum, making unmasking unsuccessful.
Thus, only elements with contributions from at least~$\SAthreshold$ clients can be correctly unmasked.
Server-client collusion resilience is addressed in Section~\ref{subsec:security-analysis}.

\subsubsection{Per-element Unmasking}
\label{subsubsec:per-element-unmasking}
To enforce per-element unmasking control, each decryptor must return masks on an element-wise basis.
A naive approach---used in Flamingo---would have decryptors send back the seeds (shares) allowing the server to regenerate the full mask efficiently.
However, this would enable the server to unmask all elements, including those that should remain hidden, thus violating (P1).

To avoid this, we have each decryptor return a mask vector instead of the seeds.
Specifically, the decryptor reveals actual mask values only for indices that meet the threshold condition.
This design ensures that the server can unmask only the intended elements, preserving per-element confidentiality.

\subsection{Efficiency Gains by Using Partial Vectors}
We reduce the size of binary indicator vectors~${\indicator}$ and the computational and communication overhead of per-element masking/unmasking mechanisms.
 

\subsubsection{Compression of Binary Indicator Vectors}
In Section~\ref{subsec:core-mechanism}, we design the protocol in which each client~$i$ sends its full binary indicator vector~$\indicator$ to the server, which then forwards~$\{\indicator\}_{i \in \clients}$ to all decryptors.
Since~$\indicator$ has the same dimension as its local update~$\inputx$ (i.e., $|\allIndices|$), sending such large vectors is communication-inefficient.
However, we can leverage the observation that local update vectors are typically sparse. 
According to~\cite{Lu2025-lw}, more than 95\% of their elements are effectively zero.
This means that~$\indicator$ is also sparse.
To reduce communication costs, we instead transmit the set~$\indices=\{k \in \allIndices \mid \indicator[k]=1\}$, which compresses the indicator information while preserving correctness.




\subsubsection{Protecting Only FC Layers}
We reduce the number of elements protected by masks by leveraging the observation that data reconstruction attacks primarily target a subset of fully connected (FC) layers, particularly those near the input or output layers~\cite{Garov2023-lu, Wang2025-vs}.
These gradients are semantically informative and less protected by architectural structures such as convolution or pooling.
Therefore, we apply our mechanism only to the parts of~$\inputx$ corresponding to selected FC layers.
As detailed in Section~\ref{subsubsec:parameter-setting}, this reduces the masking/unmasking scope to 5--40\% of the entire update vector, depending on the model architecture, and significantly lowers both computational and communication overhead.

\begin{figure}[!t]
  \centering
  \fbox{%
    \begin{minipage}{0.95\linewidth}
    \footnotesize
    \begin{phaseitem}
        \item \textbf{Setup Phase}.
        \begin{roleitem}
            \item \textbf{User~$i \in \allClients$}
            \begin{actionitem}
            \item Generates key pairs~$(g^{a_i}, a_i), (g^{b_i}, b_i)$ and stores its public keys~$(g^{a_i}, g^{b_i})$ in PKI.
            \end{actionitem}
            \item \textbf{All:}
            \begin{actionitem}
                \item Given the security parameter~$\secparam$, a public randomness~$R$, the number of clients and decryptors, the size of neighbors, threshold values~$\SAthreshold, \SSthreshold$, and the parameter $\maxDropList$.
                \item Select the decryptors~$\decryptors$ with~$R$.
            \end{actionitem}
        \end{roleitem}

        \item \textbf{Report Phase in round $\tau$.}
        \begin{roleitem}
        \item \textbf{Client $i \in \clients$:}
            \begin{actionitem}
            \item Select~$\neighbor$ with~$R$.
            \item Retrieve~$g^{a_j}$ for~$j \in \neighbor$, and~\exPart{$g^{a_u}$} and~$g^{b_u}$ for~$u \in \decryptors$ from PKI.
            \item Derive~$s_{i,j} \leftarrow g^{a_i a_j}$ for~$j \in \neighbor$, \\
                and~\exPart{$s_{i,u} \leftarrow g^{a_i a_u}$}, $k_{i,u} \leftarrow g^{b_i b_u}$ for~$u \in \decryptors$.
            \item Compute~$r_{i,j} \leftarrow \PRF(s_{i,j}, \tau)$ for $j \in \neighbor$, \\
                and~\exPart{$r_{i,u} \leftarrow \PRF(s_{i,u}, \tau)$ for $u \in \decryptors$}.
            \item \exPart{Construct~$\indicator$ and~$\indices$ from~$\inputx$, and compute as: \\
                $\inputxx = \inputx + \indicator \odot \sum_{u \in \decryptors} \PRG(r_{i,u})$.}
            \item Randomly generate~$r_i$, and mask~$\inputxx$ as:\\
                $\llbracket \inputxx \rrbracket = \inputxx + \PRG(r_i) + \sum_{j \in \neighbor} \pm \PRG(r_{i,j})$.
            \item Secret-share the seeds as~$\{\langle r_i \rangle_u\}_{u \in \decryptors} \leftarrow \SSshare(r_i, \SSthreshold, |\decryptors|)$ and~\exPart{$\{\langle r_{i,v} \rangle_u\}_{u,v \in \decryptors} \leftarrow \SSshare(\{r_{i,v}\}_{v \in \decryptors}, \SSthreshold, |\decryptors|)$}.
            \item Encrypt~$\langle r_i \rangle_u$ and~\exPart{$\{\langle r_{i,v} \rangle_u\}_{v \in \decryptors}$} with~$k_{i,u}$ for each~$u \in \decryptors$ (i.e.,~$\llbracket \langle r_i \rangle_u \rrbracket_{k_{i,u}}$ and~\exPart{$\{\llbracket \langle r_{i,v} \rangle_u \rrbracket_{k_{i,u}}\}_{v \in \decryptors}$}).
            \item Send to the server: \\
                $\llbracket \inputx \rrbracket$, \exPart{$\indices$}, $\{ \llbracket \langle r_i \rangle_u \rrbracket_{k_{i,u}} \}_{u \in \decryptors}$, \exPart{$\{\llbracket \langle r_{i,v} \rangle_u \rrbracket_{k_{i,u}} \}_{u,v \in \decryptors}$}.
            \end{actionitem}

        \item \textbf{Server:}
            \begin{actionitem}
            \item Collect messages from clients $\clients$ and aggregate~$\sum_{i \in \clients}{\llbracket \inputxx \rrbracket}$.
            \item Send to decryptor~$u \in \decryptors$: \exPart{$\{\indices\}_{i \in \clients}$} and $\{\llbracket \langle r_{i} \rangle_u \rrbracket_{k_{i,u}} \}_{i \in \clients}$.
            \end{actionitem}
        \end{roleitem}

        \item \textbf{Unmasking Phase in round $\tau$.}
        \begin{roleitem}
        \item \textbf{Decryptor $u \in \decryptors$:}
            \begin{actionitem}
            \item Retrieve~\exPart{$g^{a_i}$} and~$g^{b_i}$ for~$i \in \clients$ from the PKI.
            \item Derive~\exPart{$s_{i,u} \leftarrow g^{a_i a_u}$} and~$k_{i,u} \leftarrow g^{b_i b_u}$ for~$i \in \clients$.
            \item \exPart{Compute~$r_{i,u} \leftarrow \PRF(s_{i,u}, \tau)$ for~$i \in \clients$}.
            \item \exPart{Construct~$\clients[k]$ for each~$k$ by using~$\{\indices\}_{i \in \clients}$.}
            \item \exPart{Derive~$\submask[k]$ for each~$k$: \\
                $\submask[k]= \sum_{i \in \clients[k]}{\PRG(r_{i,u})[k]}$ if $|\clients[k]| \geq t$; otherwise, $\perp$.}
            \item Decrypt~$\{\llbracket \langle r_i \rangle_u \rrbracket_{k_{i,u}}\}_{i \in \clients}$ with~$\{k_{i,u}\}_{i \in \clients}$.
            \item Send to the server: \exPart{$\submask$} and $\{\langle r_i \rangle_u \}_{i \in \clients}$.
            \end{actionitem}

        \item \textbf{Server:}
            \begin{actionitem}
            \item Collect messages from decryptors $\decryptors_1 \subseteq \decryptors_2$.
            \item If $|\decryptors_1| \geq \SSthreshold$, reconstruct seeds: \\
                $\{r_i\}_{i \in \clients} \leftarrow \SSrecon(\{ \{ \langle r_i \rangle_u \}_{u \in \decryptors_1} \}_{i \in \clients})$; otherwise, abort.
            \item Subtract \exPart{$\sum_{u \in \decryptors_1}{\submask}$} and $\sum_{i \in \clients}{\PRG(r_i)}$ from~$\sum_{i \in \clients}{\llbracket \inputxx \rrbracket}$.
            \item \exPart{If $\decryptors_1 \neq \decryptors$, derive~$\dropList = \decryptors \setminus \decryptors_1$.}
            \item \exPart{Send to the decryptor~$u \in \decryptors_1$: $\dropList$ and~$\{\llbracket \langle r_{i,v} \rangle_u \rrbracket_{k_{i,u}} \}_{i \in \clients, v \in \dropList}$}
            \end{actionitem}
        \end{roleitem}

        \item \exPart{\textbf{Dropout Recovery Phase in round $\tau$ (if $\decryptors_1 \neq \decryptors$).}}
        \begin{roleitem}
        \item \textbf{Decryptor~$u \in \decryptors_1$:}
            \begin{actionitem}
            \item \exPart{Proceed only if~$|\dropList| \geq \maxDropList$; otherwise, abort.}
            \item \exPart{Decrypt~$\{\llbracket \langle r_{i,v} \rangle_u \rrbracket_{k_{i,u}} \}_{v \in \dropList}$. with~$\{k_{i,u}\}_{i \in \clients}$.}
            \item \exPart{Send to the server: $\{\langle r_{i,v} \rangle_u \}_{i \in \clients, v \in \dropList}$.}
            \end{actionitem}

        \item \textbf{Server:}
            \begin{actionitem}
            \item \exPart{Collect messages from decryptors~$\decryptors_2 \subseteq \decryptors_1$.}
            \item \exPart{If~$|\decryptors_2| \geq \SSthreshold$, reconstruct seeds: \\
                $\{r_{i,u}\}_{i\in \clients, u \in \decryptors \setminus \decryptors_1} \leftarrow \SSrecon(\{\langle r_{i,u} \rangle_v \}_{i \in \clients, v \in \decryptors \setminus \decryptors_1})$; \\
                otherwise, abort.}
            \item \exPart{Construct~$\{\indicator\}_{i \in \clients}$ from~$\{\indices\}_{i \in \clients}$.}
            \item \exPart{Subtract~$\sum_{i \in \clients}{\sum_{u \in \decryptors \setminus \decryptors_1}{\indicator \odot \PRG(r_{i,u})}}$ from \\
                ~$\sum_{i \in \clients}{\llbracket \inputxx \rrbracket} - \sum_{u\in \decryptors_1}{\submask}$.}
            \end{actionitem}
        \end{roleitem}
    \end{phaseitem}
    \end{minipage}
  }
  \caption{The full protocol that realizes Per-element SecAgg based on Flamingo. \exPart{Red-colored components indicate the extended procedures.} For simplicity, client dropouts are not considered.}
  \label{fig:protocol}
\end{figure}

\section{Protocol}
\label{sec:protocol}
We present the full protocol that realizes Per-element SecAgg.
It extends Flamingo by incorporating the mechanisms explained in Section~\ref{sec:framework} for (P1) and (P2), and introduces an additional phase for dropout tolerance of decryptors (P3).

\subsection{Protocol Covering for (P1) and (P2)}
\label{subsec:protocol-design}
This subsection explains the protocol phases---mainly the Report Phase and Unmasking Phase---that work together to achieve (P1) and (P2).
We focus on the red-colored parts in Fig.~\ref{fig:protocol}, which explain the extensions for Per-element SecAgg. 
For simplicity, client dropout recovery mechanisms are omitted here, but they can be achieved using Flamingo’s original techniques~\cite{Ma2023-fq} without affecting our extension.

\subsubsection{Setup Phase}
This phase is identical to the Setup Phase in Flamingo (Section~\ref{subsec:flamingo}). 
Our mechanism requires no additional setup beyond what Flamingo already provides.

\subsubsection{Report Phase}
To enable additional masking for Per-element SecAgg, each client~$i \in \clients$ first retrieves the public keys of all decryptors~$u \in \decryptors$ from the PKI and derives long-term secrets~$s_{i,u} \leftarrow g^{a_ia_u}$.
Then,~$i$ computes a round-specific seed~$r_{i,u} \leftarrow \PRF(s_{i,u}, \tau)$.

$i$ constructs an indicator vector~$\indicator$ and the set of its non-zero indices~$\indices$ from its local update~$\inputx$, following Eq.~\eqref{eq:indicator}.
Next, for per-element masking described in Section~\ref{subsubsec:per-element-masking}, $i$ masks only the non-zero elements of~$\inputx$ as follows:
\begin{equation}
    \label{eq:extra-masking}
    \inputxx = \inputx + \indicator \odot \sum_{u \in \decryptors}{\PRG(r_{i,u})},
\end{equation}
where~$\odot$ denotes element-wise multiplication.
The masked vector~$\inputxx$ is then further processed using Flamingo's original masking procedure---adding an individual mask and pairwise masks---to produce the final masked vector~$\llbracket \inputxx \rrbracket$.

To prepare for decryptor dropout recovery (explained in the next subsection), $i$ secret-shares each seed $r_{i,u}$ across all decryptors: $\{\langle r_{i,v} \rangle_u\}_{u,v \in \decryptors} \leftarrow \SSshare(\{r_{i,v}\}_{v \in \decryptors}, \SSthreshold, |\decryptors|)$.
Then, it encrypts each share~$\langle r_{i,v} \rangle_u$ using the symmetric key~$k_{i,u}$, producing~$\llbracket \langle r_{i,v} \rangle_u \rrbracket_{k_{i,u}}$.
Finally, $i$ sends the following to the server:~$\llbracket \inputxx \rrbracket$,~$\indices$, and the encrypted shares~$\{\llbracket \langle r_i \rangle_u \rrbracket_{k_{i,u}}\}_{u \in \decryptors}$ and $\{\llbracket \langle r_{i,v} \rangle_u \rrbracket_{k_{i,u}}\}_{u,v \in \decryptors}$.

Upon collecting all messages from the clients, the server aggregates the masked vectors:
\begin{equation}
    \label{eq:aggregated-masked-vector}
    \sum_{i \in \clients}{\llbracket \inputxx \rrbracket} = \sum_{i \in \clients}{\inputx} + \sum_{i \in \clients}{\PRG(r_i)} + \sum_{i \in \clients}{\sum_{u \in \decryptors}{\indicator \odot \PRG(r_{i,u})}}.
\end{equation}
To unmask Eq.~\eqref{eq:aggregated-masked-vector}, the server forwards~$\{\llbracket \langle r_i \rangle_u \rrbracket_{k_{i,u}}\}_{u \in \decryptors}$ and~$\{\indices\}_{i \in \clients}$ to each decryptor~$u \in \decryptors$.
This completes the forwarding step described in Section~\ref{subsubsec:per-element-counting}, allowing the decryptors to compute per-element contributor counts.

\subsubsection{Unmasking Phase}
\label{subsubsec:unmasking-phase}
Each decryptor~$u \in \decryptors$ first derives round-specific seeds~$\{r_{i,u}\}_{i \in \clients}$ using the public keys of clients and their long-term secrets.
Next, using the collection~$\{\indices\}_{i \in \clients}$ received from the server, $u$ constructs the set of contributors for each index~$k \in \allIndices$ as~$\clients[k] = \{i \in \clients \mid k \in \indices\}$.
This identifies the clients who reported non-zero values at index~$k$.
Based on the contributor count~$|\clients[k]|$, $u$ determines which indices satisfy the threshold~$\SAthreshold$ and generates an element-wise mask vector~$\submask$ as follows:
\begin{equation}
    \label{eq:submask}
    \submask[k]= 
    \begin{cases}
        \sum_{i \in \clients[k]}{\PRG(r_{i,u})[k]} & \text {if} \; |\clients[k]| \geq \SAthreshold \\
        \perp                                        & \text{otherwise.}
    \end{cases}
\end{equation}
where~$\perp$ denotes a null value and $\PRG(r_{i,u}[k])$ denotes the $k$-th element of the pseudorandom mask vector.
In this way, $u$ explicitly controls which indices are eligible for unmasking, in accordance with the per-element threshold policy described in Section~\ref{subsubsec:per-element-unmasking}.
Finally, the decryptor returns~$\submask$ and the decrypted seed shares~$\{\langle r_i \rangle_u\}_{i \in \clients}$ to the server.

Upon receiving messages from all decryptors, the server reconstructs the individual seeds and regenerates~$\sum_{i \in \clients}{\PRG(r_i)}$.
Then, using the sum of element-wise masks~$\sum_{u\in \decryptors}{\submask}$, the server performs the unmasking:
\begin{equation}
    \label{eq:unmasking}
    \begin{aligned}
    \boldsymbol{y} &= \sum_{i\in \clients}{\llbracket \inputxx \rrbracket} - \sum_{i\in \clients}{\PRG(r_i)} - \sum_{u\in\decryptors}{\submask} \\
    &= 
    \begin{cases}
        \sum_{i\in \clients}{\inputx[k]} & \text{if} \; |\clients[k]| \geq \SAthreshold \\
        \perp                            & \text{otherwise.}
    \end{cases}
    \end{aligned}
\end{equation}
This ensures that only elements with at least~$\SAthreshold$ contributors are unmasked, thereby fulfilling (P1).

\subsection{Protocol Covering for (P3)}
\label{subsec:decryptor-dropout-recovery}
This subsection extends the protocol presented in Section~\ref{subsec:protocol-design} to tolerate decryptor dropouts (P3).
We begin by explaining the motivation and rationale behind this extension.
We then identify a potential threat to (P2) that arises from this extension and describe our solution.
Finally, we present the detailed protocol steps.

\subsubsection{Motivation}
In the Unmasking Phase explained in Section~\ref{subsubsec:unmasking-phase}, let~$\decryptors_1 \subseteq \decryptors$ be the set of decryptors from whom the server receives element-wise masks~$\submask$.
If any decryptor drops out~(i.e., $\decryptors_1 \neq \decryptors$), the corresponding element-wise masks~$\{\submask\}_{u \in \decryptors \setminus \decryptors_1}$ are missing.
As a result, the unmasking in Eq.~\eqref{eq:unmasking} fails for all elements due to incomplete subtraction.
To satisfy (P3), we introduce a dropout recovery mechanism that is executed only when~$\decryptors_1 \neq \decryptors$.

\subsubsection{Rationale}
To tolerate decryptor dropouts, clients provide encrypted shares of all seeds in advance during the Report Phase.
If some decryptors drop out, the server sends the drop list~$\dropList=\decryptors \setminus \decryptors_1$ and corresponding ciphertexts of seed shares to the surviving decryptors~$\decryptors_1$, requests the plaintexts, and reconstructs the missing seeds.
Since only seed shares are exchanged, the communication overhead remains marginal.

This strategy may appear to contradict the per-element unmasking~(Section~\ref{subsubsec:per-element-unmasking}), which avoids disclosing seeds to the server.
However, revealing seeds for the dropped decryptors does not violate (P1), since unmasking depends on element-wise masks~$\submask$ from honest decryptors in~$\decryptors_1$, which are revealed only for elements meeting the threshold.
Thus, the server still cannot unmask under-contributed elements.


\subsubsection{Risk of Violating (P2) and Its Solution}
The dropout recovery mechanism poses a threat to (P2), in which a malicious server may add surviving and honest decryptors to the dropout list~$\dropList$, so that they are disguised as having dropped out.
The server can then obtain at least $\SSthreshold$ shares for all decryptors' seeds and fully unmask every element, violating (P1).

To counter this threat, we adopt a simple but practical strategy that requires no additional round for checking dropout decryptors.
We introduce a global constant~$\maxDropList$, which defines the maximum number of decryptor dropouts tolerated by honest decryptors.
If~$|\dropList| > \maxDropList$, the protocol is aborted by the honest decryptors.
This bound limits how many seed shares the server can receive from the honest decryptors, ensuring that at least one honest decryptor's seed cannot be recovered.
As proven in Section~\ref{subsec:security-analysis}, setting $\maxDropList = \lceil \SSthreshold / 2 \rceil$ (where $\SSthreshold$ is the threshold of the secret sharing scheme) prevents this risk while ensuring that the protocol does not abort due to decryptor dropouts.

%

\subsubsection{Procedure}
Upon detecting a decryptor dropout~(i.e., $\decryptors_1 \neq \decryptors$) during the Unmasking Phase, the server constructs a list~$\dropList = \decryptors \setminus \decryptors_1$ of the dropped decryptors.
It then sends~$\dropList$ and the encrypted seed shares~$\{\llbracket \langle r_{i,v} \rangle_u \rrbracket_{k_{i,u}}\}_{i \in \clients, v \in \dropList}$ to each surviving decryptor~$u \in \decryptors_1$.

In the Dropout Recovery Phase, each decryptor~$u \in \decryptors_1$ first verifies that~$|\dropList| \leq \maxDropList$.
Then, $u$ decrypts the received ciphertexts using the symmetric keys~$\{k_{i,u}\}_{i \in \clients}$, and returns the plaintext shares to the server.

The server collects responses from a set~$\decryptors_2$ of at least~$\SSthreshold$ decryptors and reconstructs all missing seeds~$\{r_{i,v}\}_{i\in \clients, v \in \dropList}$ via secret sharing.
Using the reconstructed seeds and the indicator vectors~$\{\indicator\}_{i \in \clients}$, the server completes the unmasking step:
\begin{equation}
\begin{aligned}
\boldsymbol{y} =\;& 
\sum_{i\in \clients}{\llbracket \inputxx \rrbracket}
\;-\; \sum_{i\in \clients}{\PRG(r_i)} \\
&\;-\; \sum_{u\in\decryptors_1}{\submask}
\;-\; \sum_{v \in \decryptors \setminus \decryptors_1}{\indicator \odot \PRG(r_{i,v})}.
\end{aligned}
\end{equation}
This yields the correct aggregate~$\boldsymbol{y}$, and achieves (P3).

\section{Cost, Security and Privacy Analysis}
\label{sec:analysis}

\begin{table}[t]
\centering
\caption{Cost per protocol phase of Our Protocol}
\resizebox{\columnwidth}{!}{%
\begin{tabular}{@{}lcc@{}}
\toprule
\textbf{}            & \textbf{Client}                & \textbf{Server}                   \\ \cmidrule(l){2-3} 
\multicolumn{1}{c}{} & \multicolumn{2}{c}{\textbf{Computation}}                           \\ \midrule
Report               & $\mathcal{O}(DK'+AK+D^3)$      & $\mathcal{O}(CK)$                 \\
Unmask               & $\mathcal{O}(CK')$             & $\mathcal{O}(DK'+CD^2+CK)$        \\
DropRcv              & $\mathcal{O}(CV)$              & $\mathcal{O}(V(CD^2+K'))$         \\ \midrule
\multicolumn{1}{c}{} & \multicolumn{2}{c}{\textbf{Communication}}                         \\ \midrule
Report               & $\mathcal{O}(K+\alpha K'+D^2)$ & $\mathcal{O}(C(D^2+DK'\alpha+K))$ \\
Unmask               & $\mathcal{O}(CK'\alpha)$       & $\mathcal{O}(D(K'+CV))$           \\
DropRcv              & $\mathcal{O}(CV)$              & $\mathcal{O}(CDV)$                \\ \bottomrule
\end{tabular}%
    \label{tab:cost}
}

\vspace{1mm}
\begin{minipage}{\columnwidth}
\footnotesize
\textbf{Notation:} 
$C$: number of clients,\;
$D$: number of decryptors,\;
$A$: size of neighbors of each client,\;
$V$: number of dropped decryptors, \;
$K$: vector size of $\inputx$, \;
$K'$: vector size to be additionally masked, \;
$\alpha$: sparsity of a vector.
\end{minipage}
\end{table}

\subsection{Computation and Communication Cost Analysis}
\label{subsec:cost-analysis}
The computation and communication costs of our protocol are summarized in Table~\ref{tab:cost}.
We define the computation cost by treating the following operations as $\mathcal{O}(1)$: one PRF evaluation, one PRG output for a vector element, one symmetric encryption/decryption, and one addition/subtraction of a vector element.
The cost of secret sharing (both $\SSshare$ and $\SSrecon$) is modeled as $\mathcal{O}(|D|^2)$ per secret.

For communication cost, we define $\mathcal{O}(1)$ as the transfer of one vector element, one item in a set, or one share.

\subsection{Security Analysis}
\label{subsec:security-analysis}
We analyze the security of our protocol.
We first show how collusion resilience (P4) is achieved by appropriate parameter settings.  
Finally, we prove that our protocol remains secure even in the presence of a malicious server, colluding users, and decryptor dropouts, thereby satisfying (P1)--(P4).

\subsubsection{Parameter Settings For (P4)}
As shown in Eq.~\eqref{eq:submask}, each decryptor returns its element-wise mask if the number of non-zero contributors at an index reaches the threshold $\SAthreshold$.
However, the element-wise mask alone cannot ensure (P1)---that the server learns only the aggregated values of elements contributed by at least $\SAthreshold$ honest clients---in the presence of colluding clients.
This is because colluding clients may maliciously flag their binary indicator vectors, making the contribution count exceed $\SAthreshold$.

To address this vulnerability, we redefine the decryptor's threshold from $t$ to $t'$ and discuss how to set it to enforce (P1) under collusion.
Clearly, $t' > \lfloor \corruptedClients |\clients| \rfloor$ is required to prevent colluding clients from meeting the threshold alone.
The goal of setting $t'$ is to hide the element values unless at least $\SAthreshold$ honest clients contribute.
We now define how to set $t'$ to ensure this.

\begin{thm}
    \label{thm:clients}
    Let $t$ be the minimum number of honest client contributions required for an element to be revealed.
    Let $t'$ be the threshold used by the decryptors to determine whether to return the element-wise mask for unmasking.
    Suppose the server may collude with up to $\lfloor \corruptedClients |\clients| \rfloor$ clients.
    Then, setting $t' = \lfloor \corruptedClients |\clients| \rfloor + \SAthreshold$ ensures that no element is unmasked unless at least $\SAthreshold$ honest clients have contributed non-zero values.
\end{thm}
\begin{sketchproof}
    Since the server can control up to $\lfloor \corruptedClients |\clients| \rfloor$ clients, it may attempt to inflate the contribution count to $\lfloor \corruptedClients |\clients| \rfloor$ at arbitrary indices.
    By setting $t' = \lfloor \corruptedClients |\clients| \rfloor + t$, the server with colluding clients cannot inflate the count to $t'$.
\end{sketchproof}

We then discuss the selection of $\SSthreshold$ and $\maxDropList$ to achieve resilience against colluding decryptors.
\begin{thm}
    \label{thm:decryptors}
    Let $\decryptors_1 \subseteq \decryptors$ be the set of online decryptors in the Unmasking Phase, and $\decryptors_2 \subseteq \decryptors_1$ be those who respond in the Dropout Recovery Phase.
    Suppose the server may collude with up to $\lfloor \corruptedDecryptors |\decryptors| \rfloor$ decryptors and up to $\lfloor \dropDecryptors |\decryptors| \rfloor$ honest decryptors may drop out, under the constraint $\dropDecryptors + \corruptedDecryptors < 1/3$.
    Then, setting $\SSthreshold = \lfloor 2|\decryptors|/3 \rfloor+1$, $\maxDropList = \lceil \SSthreshold / 2 \rceil$ simultaneously ensures the following: \\
    (Recovery) The server can reconstruct the seeds of all dropped decryptors~$v \in \decryptors \setminus \decryptors_1$ if $|\decryptors_2| \geq \SSthreshold$.\\
    (Security) A malicious server cannot reconstruct the complete seeds of all ($\lfloor (1 - \corruptedDecryptors) |\decryptors| \rfloor $) honest decryptors.
\end{thm}
\begin{sketchproof}
    Each decryptor $u \in \decryptors_2$ returns $|\clients|$ seed shares for each dropped decryptor $v \in \dropList$ reported by the server.  
    Treating this response as one unit per $v$, the total number of shares the server receives is $ \maxDropList \cdot \lfloor (1 - \dropDecryptors - \corruptedDecryptors) |\decryptors| \rfloor $.

    \textit{(Recovery)}  
    To reconstruct the seeds of $ \lfloor \dropDecryptors |\decryptors| \rfloor $ dropped decryptors, the server requires $ \lfloor \dropDecryptors |\decryptors| \rfloor \cdot (\SSthreshold - \lfloor \corruptedDecryptors |\decryptors| \rfloor) $ shares.
    The server can successfully recover the missing seeds if the number of received shares is at least this amount, i.e.,  
    $ \maxDropList \cdot \lfloor (1 - \dropDecryptors - \corruptedDecryptors) |\decryptors| \rfloor \geq \lfloor \dropDecryptors |\decryptors| \rfloor \cdot (\SSthreshold - \lfloor \corruptedDecryptors |\decryptors| \rfloor) $.
    In the worst-case setting for recovery, where $ \dropDecryptors \to 1/3 $ and $ \corruptedDecryptors \to 0 $, this simplifies to $ \maxDropList \geq \SSthreshold / 2 $.

    \textit{(Security)}  
    To reconstruct all seeds of $ \lfloor (1 - \corruptedDecryptors) |\decryptors| \rfloor $ honest decryptors, a malicious server requires $ \lfloor (1 - \corruptedDecryptors) |\decryptors| \rfloor \cdot (\SSthreshold - \lfloor \corruptedDecryptors |\decryptors| \rfloor) $ shares.
    To ensure security, the number of shares the server obtains must be strictly less than this quantity:  
    $ \maxDropList \cdot \lfloor (1 - \dropDecryptors - \corruptedDecryptors) |\decryptors| \rfloor < \lfloor (1 - \corruptedDecryptors) |\decryptors| \rfloor \cdot (\SSthreshold - \lfloor \corruptedDecryptors |\decryptors| \rfloor) $.
    In the worst-case for security, where $ \dropDecryptors \to 0 $ and $ \corruptedDecryptors \to 1/3 $, this simplifies to $ \maxDropList < \SSthreshold - |\decryptors| / 3 $.

    From both conditions, $ \maxDropList $ is $ \SSthreshold / 2 \leq \maxDropList < \SSthreshold - |\decryptors| / 3 $ is derived.
    This range is non-empty if $ \SSthreshold > 2|\decryptors| / 3 $.  
    Accordingly, the optimal parameters that satisfy both recovery and security are given by $ \SSthreshold = \lfloor 2|\decryptors| / 3 \rfloor + 1 $ and $ \maxDropList = \lceil \SSthreshold / 2 \rceil $.
\end{sketchproof}

\subsubsection{Per-element Threshold Aggregation}
We now show that our protocol ensures per-element threshold aggregation.
\begin{thm}
    Let the parameters be~$t' = \lfloor \corruptedClients |\clients| \rfloor + t$, $\SSthreshold = \lceil 2|\decryptors|/3 \rceil + 1$, and $\maxDropList=\lfloor \SSthreshold/2 \rfloor$.
    Suppose the server colludes with up to~$\lfloor \corruptedClients |\clients| \rfloor$ clients and $\lfloor \corruptedDecryptors |\decryptors| \rfloor$ decryptors, and up to $\lfloor \dropDecryptors |\decryptors| \rfloor$ decryptors may drop out, with $\dropDecryptors + \corruptedDecryptors < 1/3$.
    Assume the underlying cryptographic primitives, PRG-based masking encryption~\cite{Castelluccia2009-zk} and the Flamingo protocol are secure.
    Then, the server learns only the sum of elements contributed by at least $\SAthreshold$ honest clients; elements with fewer contributions remain hidden.
\end{thm}
\begin{sketchproof}
    A full proof via a hybrid argument is omitted; we sketch the key ideas.
    By the security of Flamingo, the server receives only the masked sum~$\sum_{i \in \clients}{\inputxx}$.
    This holds even if the server is semi-honest or malicious.

    If the number of contributors at an index is at least $\SAthreshold$, then decryptors return the corresponding element-wise masks.
    These are sufficient to cancel out all per-element masking, allowing the server to recover the true sum at that index.
    For indices with fewer than $\SAthreshold$ contributors, at least one honest decryptor will withhold its mask.
    Due to the unpredictability of PRG outputs and the security of the underlying key agreement, this missing mask renders the masked value computationally indistinguishable from random.

    Theorems~\ref{thm:clients} and~\ref{thm:decryptors} guarantee that the server cannot forge contribution counts or manipulate dropout reports to gain access to missing masks.
    Thus, elements are revealed if and only if at least $\SAthreshold$ clients contribute non-zero values.
\end{sketchproof}

\begin{figure*}[t]
   \centering
   \includegraphics[width=\linewidth]{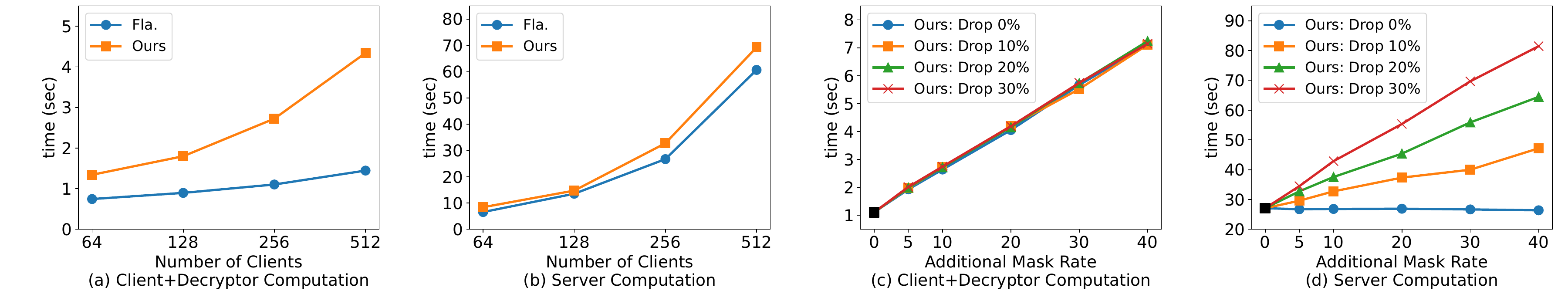}
    \caption{Computation time vs. number of clients and masking ratio (model update vector dimension is 5M, number of decryptors is 40).}
 \label{fig:eval-comp}
\end{figure*}

\begin{figure*}[t]
   \centering
   \includegraphics[width=\linewidth]{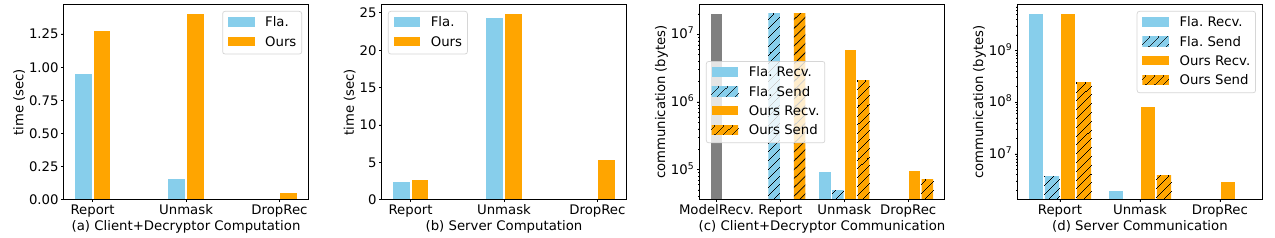}
    \caption{Breakdown of protocol overhead in each phase (model update vector dimension is 5M, number of clients is 256, number of decryptors is 40, and dropout rate of decryptors is 10\%).}
 \label{fig:eval-phase}
\end{figure*}

\subsection{Privacy Analysis}
\label{subsec:privacy-analysis}
Our protocol reveals to the server and decryptors the indices at which each client has non-zero updates, via the set of indices~$\indices$.
This subsection discusses the potential privacy implications of this design choice.

One concern is that index exposure may lead to data reconstruction.
Us Sami et al.~\cite{Us-Sami2024-bm} argue that, under SecAgg, knowing the indices of non-zero updates can allow the server to reconstruct clients' datasets by solving linear systems.
However, this attack assumes the server sends an unchanged global model to the same client for hundreds of rounds, which is unrealistic in our setting, where clients are randomly reselected in each round (see Section~\ref{subsec:system-model}).

Another possible risk is attribute inference.
Pasquini et al.~\cite{Pasquini2022-sb} demonstrate that, by observing which indices are non-zero, the server can infer sensitive properties of a client's dataset.
As noted in Section~\ref{subsec:design-goals}, our framework specifically targets data reconstruction attacks and does not protect against such inference.
Addressing this remains future work.

\section{Evaluation}
\label{sec:evaluation}
We compare our proposed protocol with Flamingo and evaluate the additional overhead.
We also examine the impact of Per-element SecAgg on model performance.

\subsection{Experimental Setup}
\label{subsec:experimental-setup}
\subsubsection{Implementation}
We implement our protocol on the ABIDES simulator~\cite{Byrd2020-ai}, which has been used to evaluate several SecAgg protocols~\cite{Ma2023-fq, Guan2025-jt, Karthikeyan2024-sm}.
Cryptographic primitives are implemented by using the same modules used in Flamingo.

\subsubsection{Parameter Settings}
\label{subsubsec:parameter-setting}
To reflect practical FL scenarios in which a malicious server may launch data reconstruction attacks under SecAgg, we configure the parameters as follows.

\noindent
\textbf{Model Update Vector Dimensionality. }
We assume a cross-device setting in which clients are resource-constrained user devices such as smartphones.
In such settings, models typically contain several million parameters~(e.g., ResNet-9 has 4.9M, LSTM has 8.3M~\cite{Dorfman2023-gt}, and small Transformer models around 4.1M~\cite{Ro2022-iw}).
Based on this observation, we fix the dimensionality of the model update vector to 5M for all evaluations.

\noindent
\textbf{Additional Mask Rate}
As shown in Section~\ref{subsec:cost-analysis}, the overhead of our Per-element SecAgg depends on the proportion of the update vector that is masked.
When data reconstruction attacks rely solely on model parameter manipulation, the masked portion typically remains below 10\% for architectures such as ResNet and Transformer (e.g., Wen et al.\cite{Wen2022-rw} propose an attack that observes only the output layer, which accounts for 4.4\% of the parameters in ResNet-18~\cite{Strassenburg2022-vv}).
However, attacks that involve modifications to the model architecture---such as inserting additional linear layers for reconstruction~\cite{Zhao2024-wy}---may require a higher masking rate.
This also holds for potential future attacks that exploit deeper model structures.
Therefore, we fix the update vector size to 5M and vary the masking ratio from 5\% to 40\% in our evaluation.

\noindent
\textbf{Sparsity of update vectors. }
To prevent reconstruction attacks, elements in the model update vector with magnitudes below a threshold~$\lambda$ should be treated as zeros.
This is aligned with threshold-based sparsification approaches in FL~\cite{Lu2025-lw, Lu2025-cc}.
According to~\cite{Lu2025-lw}, setting~$\lambda$ between ${10}^{-3}$ and~${10}^{-2}$ results in over 99\% of elements being zeroed out.
In our evaluation, we conservatively assume a sparsity level of 95\%.

\subsection{Computation Time and Communication Overhead}
\subsubsection{When Additional Mask Rate is 10\%}
We evaluate the computation time and communication overhead in the most practical setting, where the additional mask rate is 10\%.
Figs.~\ref{fig:eval-comp}a and \ref{fig:eval-comp}b show the computation time for a user (client + decryptor) and the server, respectively, as the number of clients ($|\clients|$) increases, in comparison with Flamingo (``Fla.") and our protocol with Per-element SecAgg (``Ours").

At 256 clients, the additional time compared to Flamingo is 1.6s on the user side and 6.0s on the server side, totaling just 7.6s per round. 
While both increase with $|\clients|$, the overhead remains modest.
Fig.~\ref{fig:eval-phase} breaks down the computation time and communication overhead at 256 clients.
On the user side, 1.2s (75\%) of the overhead comes from the decryptor’s Unmasking phase (Fig.~\ref{fig:eval-phase}a), which is acceptable, as decryptors do not perform training and remain lightweight.
On the server side, 5.3s (80\%) stems from Dropout Recovery (Fig.~\ref{fig:eval-phase}b).

Figs.~\ref{fig:eval-phase}c and \ref{fig:eval-phase}d show the communication overheads, with increases of 1.21× on the user side and 1.07× on the server side.
User-side growth is more notable.
This is mainly due to the decryptor’s reception of the index sets $\{\indices\}_{i \in \clients}$ (“Recv.” in Unmask) and the transmission of element-wise masks (“Send” in Unmask), which together account for 96\% of the increase.
Nevertheless, this is acceptable, as both are much lighter than client communication, such as downloading the global model (“Model Recv.”) or uploading updates (“Send” in Report).

\subsubsection{Computation Time under Varying Additional Mask Rate}
Figs.~\ref{fig:eval-comp}c and \ref{fig:eval-comp}d show the computation time for a user and the server, respectively, as the additional mask rate varies.
Note that the case of 0\% corresponds to the original Flamingo.
We also vary the decryptor dropout rate across 0\%--30\%.

On the user side, computation time increases roughly linearly with the additional mask rate.
The maximum overhead is a 6.4× increase over Flamingo when 40\% of elements are additionally masked.
The user time is largely unaffected by decryptor dropout, since the corresponding recovery involves only lightweight symmetric decryption.

The server-side computation time also grows linearly with the additional mask rate, and the slope increases with higher dropout rates.
The highest overhead---2.9× over Flamingo---occurs at 40\% additional masking and 30\% dropout.
Notably, when the dropout rate is 0\%, the server time remains nearly identical to Flamingo, regardless of the additional mask rate.
This is because, in the absence of dropout, the server only performs element-wise subtraction on masks, a lightweight operation as shown in Fig.~\ref{fig:eval-phase}b.

\begin{figure}[t]
   \centering
   \includegraphics[width=0.95\linewidth]{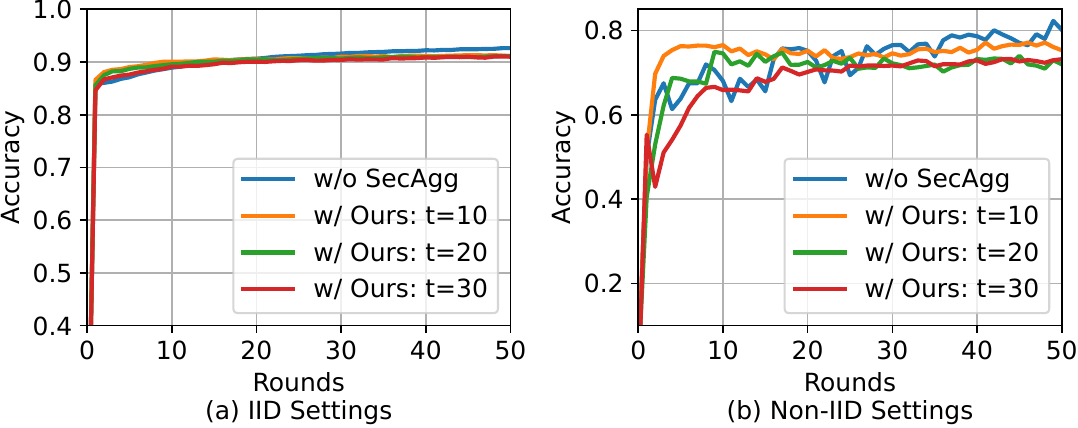}
    \caption{Accuracy under MNIST dataset (number of clients is 100, the optimizer is FedAvg, number of local training epochs in each client is 5).}
 \label{fig:accuracy-test}
\end{figure}

\subsection{Impact on Model Performance}
In Per-element SecAgg, elements with fewer than $\SAthreshold$ non-zero contributions in a round are not revealed to the server, and the corresponding elements in the global model cannot be updated.
While this design enhances privacy—especially with larger $\SAthreshold$, it may affect model accuracy.

We evaluate this effect using a three-layer fully connected neural network and the MNIST dataset~\cite{Guan2025-jt}, with 100 clients.
Model updates are sparsified by thresholding small-magnitude elements, yielding about 95\% sparsity.
In the IID setting, each client holds data uniformly sampled across all 10 labels. 
In the non-IID setting, each client holds data from 2 labels.

Under the IID setting (Fig.~\ref{fig:accuracy-test}a), Per-element SecAgg shows almost the same convergence speed and final accuracy as the baseline (``w/o SecAgg''), even when $\SAthreshold$ is increased; the final accuracy difference remains as small as 0.01.
Under the non-IID setting (Fig.~\ref{fig:accuracy-test}b), the accuracy transition becomes less stable due to data heterogeneity, and when $\SAthreshold = 20, 30$, the final accuracy degrades by up to 0.05 compared to the baseline.

This is likely due to reduced overlap of non-zero indices.
As noted in~\cite{Qiu2021-hj, Guastella2024-nb}, IID clients are more likely to update the same elements, making threshold satisfaction easier—even when the updates are highly sparse.
In contrast, under non-IID settings, heterogeneous data distributions lead to diverging update patterns among clients.
This divergence, combined with high update sparsity, makes the overlap of non-zero indices especially low, which in turn significantly hinders the server’s ability to collect at least $\SAthreshold$ contributions at each index.
Improving robustness in non-IID settings under high $\SAthreshold$ remains an important direction for future work.

\section{Related Work}
\label{sec:related}
This section summarizes data reconstruction attacks that exploit the sparsity of model update vectors, which our Per-element SecAgg can potentially prevent.
Existing countermeasures and their limitations are already discussed in Section~\ref{sec:introduction}.

A common strategy exploits the ReLU activation, which outputs zero for negative inputs.
A malicious server can manipulate model weights so that only a specific sample yields a non-zero output.
When distributed to a client, the resulting update reveals gradient information corresponding to that sample, as demonstrated by Pasquini et al.\cite{Pasquini2022-sb} and Boenisch et al.\cite{Boenisch2023-ow}.
Other approaches aim for more accurate reconstruction by modifying not only model parameters but also the architecture itself.
Fowl et al.\cite{Fowl2022-an} and Zhao et al.\cite{Zhao2024-wy} propose attacks that insert linear layers with ReLU activation into the global model to directly extract data from the corresponding updates.
Wen et al.~\cite{Wen2022-rw} amplify gradients only for target classes or features by carefully altering weights and biases.
Since this attack isolates updates by feature rather than by client, distributing the same manipulated model to all clients can still enable data extraction, effectively bypassing SecAgg.
Although most existing attacks target image reconstruction, Chu et al.~\cite{Chu2022-bu} extend this to text.
They focus on Transformer models and amplify updates associated with specific keywords by modifying multi-head self-attention weights, enabling the reconstruction of inputs such as credit card numbers.




\section{Conclusion}
\label{sec:conclusion}
This paper proposes Per-element SecAgg, a new mechanism that enhances SecAgg in FL by revealing aggregated values only when at least $\SAthreshold$ clients contribute non-zero elements.
We design a concrete protocol by integrating this mechanism into the Flamingo SecAgg protocol without introducing new cryptographic primitives.
We evaluate the additional overhead and model performance impact of Per-element SecAgg, demonstrating its practicality in realistic settings.



\bibliographystyle{IEEEtran}
\bibliography{IEEEfull, definition, reference.bib}

\end{document}